\documentclass[epj,nopacs]{svjour}
\usepackage[T1]{fontenc}
\usepackage[latin9]{inputenc}
\usepackage{color}
\usepackage{array}
\usepackage{amstext}
\usepackage{graphicx}
\usepackage{esint}
\usepackage{graphics}
\begin{document}
\title{Global analysis of fragmentation functions and their application to polarized top 
quark decays considering new {\it BABAR} and {\it Belle} experimental data}

\author{S. M. Moosavi Nejad\inst{1,3}, M. Soleymaninia\inst{3}, A. N. Khorramian\inst{2,3}}

\institute{Faculty of Physics, Yazd University, P.O. Box
89195-741, Yazd, Iran \and
Faculty of Physics, Semnan University, 35131-19111 Semnan, Iran
\and
School of Particles and Accelerators,
Institute for Research in Fundamental
Sciences, P.O.Box
19395-5531, Tehran, Iran
}

\date{\today}
%
\abstract{
Recently, the {\it Belle}  and {\it BABAR} Collaborations  published the
single-inclusive electron-positron annihilation data
at the center of mass energies ($\sqrt{s}$) of $10.52$~GeV and $10.54$~GeV, respectively.
These new data offer one the possibility to determine the nonperturbative initial conditions of
fragmentation functions much more accurately. In our previous work \cite{Soleymaninia:2013cxa}, we 
extracted the fragmentation functions for $\pi^\pm$ 
and $K^\pm$ particles at next-to-leading order (NLO) including these new data, for the first time.
These new data are in the regions of larger scaled-energy $z$ and lower $\sqrt{s}$.
Our main purpose is to show that adding  
these new data in our analysis how much improve the fragmentation functions of $\pi^\pm$ 
and $K^\pm$ at NLO. We hope this analysis can obvious the effects of these recent data on FFs 
for whom is studying the behavior of light meson FFs.
We also apply, for the first time, the extracted fragmentation functions to make our predictions for the 
scaled-energy distributions of $\pi^\pm$ 
and $K^\pm$ inclusively produced in polarized top quark decays at NLO.}

\maketitle

\section{Introduction}
\label{sec1}

Fragmentation functions $D_i^h(z, \mu_f^2)$, which can be described as the probability for
a parton $\it{i}$ at the factorization scale $\mu_f$ to fragment into a hadron $\it{h}$
carrying away a fraction $\it{z}$ of its momentum, are 
the key quantities  for calculating the hadron production cross section,  investigating the properties of
quarks in heavy ion collisions and spin physics. Specially, to study the properties of top quark at 
LHC, one of the proposed channels is to consider the energy spectrum of outgoing mesons from top decays, 
in which by having parton-level differential decay rates \cite{Kniehl:2012mn,Cacciari:2002re,Corcella:2001hz} 
and the fragmentation functions (FFs) of partons into hadrons, one can calculate the energy distribution of observed mesons.\\
Generally, there are two main approaches to evaluate the FFs. 
The first approach is based on the fact that the FFs for mesons containing a heavy quark can be computed
theoretically using perturbative QCD (pQCD) \cite{Ma:1997yq,Braaten:1993rw,Chang:1991bp,Braaten:1993mp,Scott:1978nz}.
The first theoretical attempt to explain the procedure of hadron
production from a heavy quark was made by Bjorken \cite{Bjorken:1977md} by using a naive quark-parton model (QPM). He deduced that
the inclusive distribution of heavy hadron should peak almost at $z = 1$, where z refers to the scaled-energy variable.
The pQCD scheme was followed by Peterson \cite{Peterson:1982ak}, Suzuki \cite{Suzuki:1977km}, 
Amiri and Ji \cite{Amiri:1986zv}, while in this scheme 
Suzuki calculates the heavy FFs using a convenient Feynman diagram. One of us, using the Suzuki approach has calculated
the FF for c-quark to split into S-wave $D^0/D^+$ meson \cite{Nejad:2013vsa} and the initial FF of
gluon to split into S-wave charmonium state ($J/\psi$) \cite{Nejad:2014iba} to leading order in the QCD coupling constant $\alpha_s$.\\
In the second approach, which is frequently used to obtain the
FFs, these functions are extracted from experimental data analysis using the data from 
$e^+e^-$  reactions, lepton-hadron and hadron-hadron scattering processes.
This situation is very similar to the determination of the parton distribution functions (PDFs).
Among all scattering processes, the
best processes which provide a clean environment to determine the FFs are $e^+e^-$ annihilation
processes \cite{Leitgab:2013qh,Lees:2013rqd}.
There are several theoretical studies on QCD analysis
of FFs which used special parametrization and different experimental data in their global analysis.
Recent extracted FFs are related to SKMA \cite{Soleymaninia:2013cxa}, AKK \cite{Albino:2008fy}, 
DSSV \cite{deFlorian:2007aj} and HKNS \cite{Hirai:2007cx}, using different phenomenological models.
Since the hadronization mechanism is universal and independent of the perturbative process
which produces partons, one can exploit, for example, the existing data on $e^+e^-\rightarrow b\bar b$
events to fit such models and describe the b-quark non-perturbative fragmentation in
other processes, such as top decay.\\
In our previous work \cite{Soleymaninia:2013cxa}, we determined the nonperturbative $\pi^\pm$ and $K^\pm$ FFs,
both at Leading Order (LO) and NLO
in the modified Minimal-Subtraction ($\overline{MS}$) factorization scheme, by global fitting the fractional-energy
spectra of these hadrons obtained from the single-inclusive $e^+e^-$ annihilation (SIA) and the
semi-inclusive deep inelastic scattering (SIDIS) data from HERMES and COMPASS.
However, data for the production of $\pi^\pm$ is generally more accurate  than for the production of 
other particles due to the high affluence of $\pi^\pm$ in the existing particle sample.
New data on $\pi^\pm/K^\pm$ production with much higher accuracy at larger $z$ and lower $\sqrt{s}$
have been presented
by the {\it Belle} Collaboration  at $\sqrt s = 10.52$~GeV \cite{Leitgab:2013qh} and {\it BABAR} 
Collaboration  at $\sqrt s = 10.54$~GeV \cite{Lees:2013rqd}. 
These new data offer one the possibility to determine the nonperturbative initial conditions of
the FFs much more accurately. Note that
large $z$, low $\sqrt{s}$ data impose more constraints on the gluon fragmentation function than the smaller $z$, higher $\sqrt{s}$ ones do.
Furthermore, the large span in center-of-mass (c.m.) energy ($\sqrt{s}$)
ranging from $10.52$~GeV way up to $91.2$~GeV 
\cite{aleph91,delphi91,delphi91-2,opal91,sld91,tasso34_44,tpc29,topaz58} provides us with a powerful lever
arm to test the Dokshitzer-Gribov-Lipatov-Altarelli-Parisi (DGLAP) \cite{dglap} evolution of the FFs.

In the Standard Model (SM) of particle physics top quark has short
lifetime and it decays before hadronization takes place, then its full polarization content is retained
when it decays. Therefore we can study the top spin state using the angular distributions 
of its decay products \cite{Nejad:2013fba}. In this work we make our predictions for the
scaled-energy distribution of $\pi^\pm/K^\pm$ inclusively produced in polarized
top quark decays, $t(\uparrow)\rightarrow W^++b(\rightarrow \pi^\pm/K^\pm+X)$, using the extracted FFs from new data. 
In particular, these predictions will enable us to deepen our understanding of the nonperturbative aspects
of $\pi^\pm/K^\pm$ formation by hadronization and to pin down the $b\rightarrow \pi^\pm/K^\pm$ FFs.\\
This paper is organized as follows. In section~\ref{sec2} we describe our formalism and
parametrization form for pion and kaon fragmentation densities. In section~\ref{sec3} we explain 
the effect of {\it Belle} and {\it BABAR} data on FFs determination. Our predictions for the energy spectrum 
of pions and kaons produced in polarized top quark decays are presented in section~\ref{sec4}.
In section~\ref{sec5} global minimization and error calculation method are described.
Our conclusion is summarized in section~\ref{sec6}.

\section{QCD analysis of fragmentation functions}
\label{sec2}
The FFs are the nonperturbative part of the hadronization processes
and they have an important role in the calculation of single-inclusive hadron production in any reaction.
According to the factorization theorem, the leading twist component of any single hadron inclusive production measurement
can be remarked as the convolution of fragmentation functions with the equivalent
productions of real partons, which are perturbatively calculable, up to possible PDFs  
to account for any hadrons in the initial state. As an example, the production of hadron $H$ in the typical 
scattering process of $A+B\rightarrow H+X$, can be expressed as
\begin{eqnarray}
d\sigma&=&\sum_{a,b,c}\int_0^1 dx_a\int_0^1 dx_b\int_0^1 dz f_{a/A}(x_a, \mu)f_{b/B}(x_b, \mu)\times\nonumber\\
&&d\hat\sigma(a+b\rightarrow c+X)D_{c\rightarrow H}(z, \mu),
\end{eqnarray}
where $\mu$ is a factorization scale, $a$ and $b$ are incident partons in the colliding initial hadrons $A$ and $B$ respectively,
$f_{a/A}$ and $f_{b/B}$ are the PDFs at the scale $\mu$, $c$ is the
fragmenting parton (either a gluon or a quark) and $X$ stands for the unobserved jets. 
Here, $D_{c\rightarrow H}(z, \mu)$ is the fragmentation function at the scale $\mu$ which
can be obtained by evolving from the initial FF $D_{c\rightarrow H}(z, \mu_0)$
using the DGLAP renormalization group equations \cite{dglap}.\\
One of the most current approaches to determine the FFs is based on the
single-inclusive hadron production data
through $e^-e^+$ annihilation.  From the factorization theorem, 
the cross section can be expressed in terms of the partonic hard scattering cross
 sections and the nonperturbative FFs $D_i^H(z, Q^2)$, i.e.
\begin{eqnarray}\label{fac}
\frac{1}{\sigma_{tot}}\frac{d}{dz}\sigma(e^+e^-\rightarrow HX)=\sum_i C_i(z, \alpha_s) \otimes D_i^H(z, Q^2),\nonumber\\
\end{eqnarray}
where, the function $D_i^H(z, Q^2)$ indicates the probability to find the
hadron $H$ from a parton $i(=g, u, d, s, \cdots)$ with the energy fraction $z=2E_H/\sqrt{s}$
and $C_i(z, \alpha_s) $ is the Wilson coefficient function based on the partonic cross section $e^+e^-\rightarrow q\bar q$
which is calculated in the perturbative QCD \cite{Kretzer:2000yf,Kniehl:2005de,Binnewies:1994ju}, and the convolution
integral is defined as $f(z)\otimes g(z)=\int_z^1 dy/y f(y)g(z/y)$. In the equation above,
 $X$ stands for the unobserved jets and  $\sigma_{tot}$  is the total hadronic
cross section \cite{Kneesch:2007ey}.\\
There are several different strategies to extract the FFs from data analysis so in the present analysis we adapt
the zero-mass variable-flavor-number (ZM-VFN) scheme \cite{Kneesch:2007ey}.
This scheme works best for high energy scales, where the mass of
heavy quarks are set to zero from the start and the non-zero values of the c- and b-quark
masses only enter through the initial conditions of the FFs, and the mass of the heavy hadron sets
the lower bound on the scaling variable $z$.
In the phenomenological approach, the FFs are parameterized in a convenient functional form at the initial scale $\mu_0$
in each order, i.e. LO and NLO. Various phenomenological models like Peterson
model \cite{Peterson:1982ak}, Power model \cite{Kniehl:2008zza},
Cascade model \cite{Webber:1983if} etc., have been developed to describe the FFs.
Here, we apply very flexible parameterization form for the $\pi$ and $K$ FFs at NLO, considering 
SIA data from LEP ({\it ALEPH} \cite{aleph91}, {\it DELPHI} \cite{delphi91,delphi91-2} and {\it OPAL} \cite{opal91} Collaborations), 
SLAC ({\it BABAR} \cite{Lees:2013rqd}, {\it SLD} \cite{sld91} and {\it TPC} \cite{tpc29} Collaborations), DESY ({\it TASSO} \cite{tasso34_44}
Collaboration) and KEK ({\it Belle} \cite{Leitgab:2013qh} and {\it TOPAZ} \cite{topaz58} Collaborations) and SIDIS data
from HERMES05 \cite{Hermes05} and COMPASS \cite{Alekseev:2009ac,Alekseev:2010ub}. At the initial scale $\mu_{0}$
this parametrization contains a functional form as
\begin{equation}
\label{ff-input}
D_i^H(z,\mu_{0}^{2}) =
N_i z^{\alpha_i}(1-z)^{\beta_i} [1-e^{-\gamma_i z}],
\end{equation}
which  is an appropriate form for the light hadrons. To control
medium $z$ region and to improve the accuracy of the global fit the term $[1-e^{-\gamma_i z}]$ is considered.
The free parameters $N_i$, $\alpha_i$, $\beta_i$ and $\gamma_i$ are determined by global fitting $\chi^{2}$
using the SIA and SIDIS data and their $\mu$ evolution is determined
by the DGLAP  equations. Our results are listed 
in Tables.~\ref{tab:nlopionpara} and \ref{tab:nlokaonpara} for $\pi$ and $K$ FFs.
The initial scale $\mu_{0}$ is different for partons so that the value of $\mu_{0}^{2}=1$~GeV$^2$ 
is chosen for splitting of the light-quarks $(u, d, s)$  and gluon into the $\pi^\pm/K^\pm$-mesons
and for the $c-$ and $b-$quarks it is taken to be $\mu_{0}^{2}=m_{c}^{2}$ and $\mu_{0}^{2}=m_{b}^{2}$, respectively.\\
According to the partonic structure of $\pi^- (\bar{u}d)$ and $K^- (\bar{u}s)$, the following assumptions
are considered during our calculations
\begin{eqnarray}
\label{minus1}
D_{i}^{\pi^{-}}(z,\mu_{0}^{2}) = D_{\bar{i}}^{\pi^{+}}(z,\mu_{0}^{2}),\\ \nonumber
D_{i}^{K^{-}}(z,\mu_{0}^{2}) = D_{\bar{i}}^{K^{+}} (z,\mu_{0}^{2}),
\end{eqnarray}
where $i=u,d,s,c,b$ and for the gluon FFs, it reads
\begin{eqnarray}
\label{minus2}
D_{g}^{\pi^{-}}(z,\mu_{0}^{2}) = D_{g}^{\pi^{+}}(z,\mu_{0}^{2}),\\ \nonumber
D_{g}^{K^{-}}(z,\mu_{0}^{2}) = D_{g}^{K^{+}}(z,\mu_{0}^{2}).
\end{eqnarray}

%
\begin{table*}[th]
\caption{\label{tab:nlopionpara} Values of fit parameters for the  $\pi^+$ meson at NLO
  in the starting scales.}
\centering
\hspace{1cm}
\begin{tabular}{cccccc}
\hline
\hline
flavor $i$ &$N_i$ & $\alpha_i$ & $\beta_i$ &$\gamma_i$\\
\hline
$u, \overline{d}$ &$ 1.049\pm0.563$&$-1.916\pm0.421$& $0.977\pm0.304$&$0.964\pm0.65$
\\
$\overline{u}, d, s, \overline{s}$ & $9.968\pm7.441$&$-0.516\pm0.481$&$5.952\pm1.565$&$1.898\pm1.885$
\\
$c, \overline{c}$ &$ 0.946\pm0.859$&$-1.723\pm0.451$& $3.590\pm1.280$&$1.947\pm1.981$
\\
$b, \overline{b}$ & $0.869\pm0.550$&$-2.059\pm0.234$&$ 5.803\pm1.460$& $1.561\pm1.054$
\\
$g$              & $219.507\pm44.789$&$ 1.073\pm0.362$& $7.505\pm1.140$&$2.142\pm1.411$
\\
\hline
\hline
\end{tabular}
\end{table*}

\begin{table*}[th]
\caption{\label{tab:nlokaonpara} Values of fit parameters for the $K^+$ meson at NLO
 in the starting scales.}
\centering
\hspace{1cm}
\begin{tabular}{cccccc}
\hline
\hline
flavor $i$ &$N_i$ & $\alpha_i$ & $\beta_i$ &$\gamma_i$\\
\hline
$u$ & $0.660\pm0.156$&$-1.584\pm 0.342$& $0.858\pm0.227$&$0.390\pm0.107$
\\
$\overline{s}$ & $17.769\pm7.775$&$0.708\pm0.390$&$2.479\pm0.316$&$0.665\pm0.218$
\\
$\overline{u}, d, \overline{d}, s$ &$ 6.467\pm1.587$&$0.028\pm0.547$&$ 7.338\pm0.819$&$3.299\pm1.282$
\\
$c, \overline{c}$ & $7.217\pm1.013$&$0.550\pm0.113$&$ 5.366\pm0.314$&$-$
\\
$b, \overline{b}$ &$ 14.675\pm3.227$&$0.293\pm0.080$&$ 10.882\pm0.943$&$-$
\\
$g$              & $2.383\pm0.381$&$5.714\pm0.696$&$ 0.892\pm0.085$& $53542.030\pm5.859$
\\
\hline
\hline
\end{tabular}
\end{table*}
\begin{figure*}[th]
\centerline{\includegraphics[width=0.7\textwidth,angle=-90]{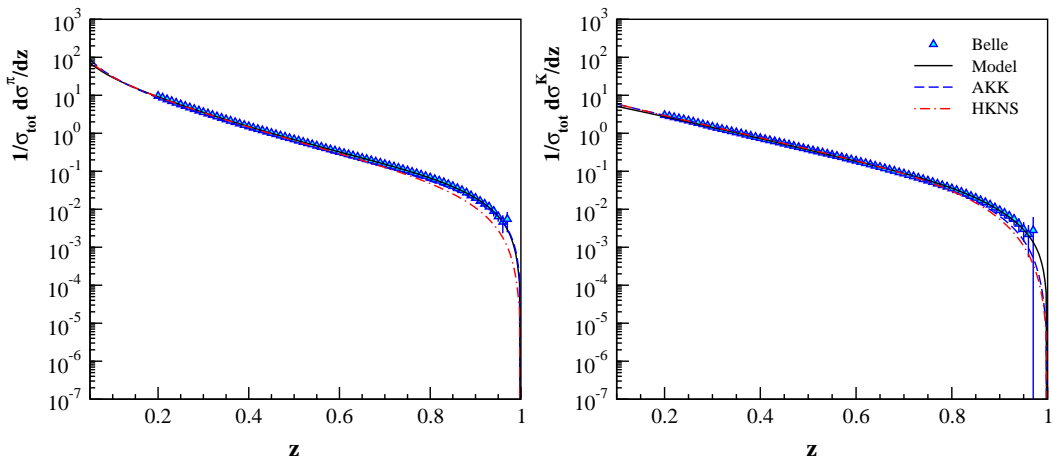}}
\vspace{-3cm}
\caption{Comparison of our NLO results for $1/\sigma_{tot}\times d\sigma^{i}/dz~ (i=\pi, K)$ 
with pion and kaon SIA data from {\it Belle} at $Q=10.52$~GeV~\cite{Leitgab:2013qh}. Our model (solid lines) 
is also compared with AKK (dashed lines) \cite{Albino:2008fy}  and HKNS (dot-dashed lines) \cite{Hirai:2007cx}.}
\label{Belledata}
\end{figure*}
%
\begin{figure*}[th]
\centerline{\includegraphics[width=0.7\textwidth,angle=-90]{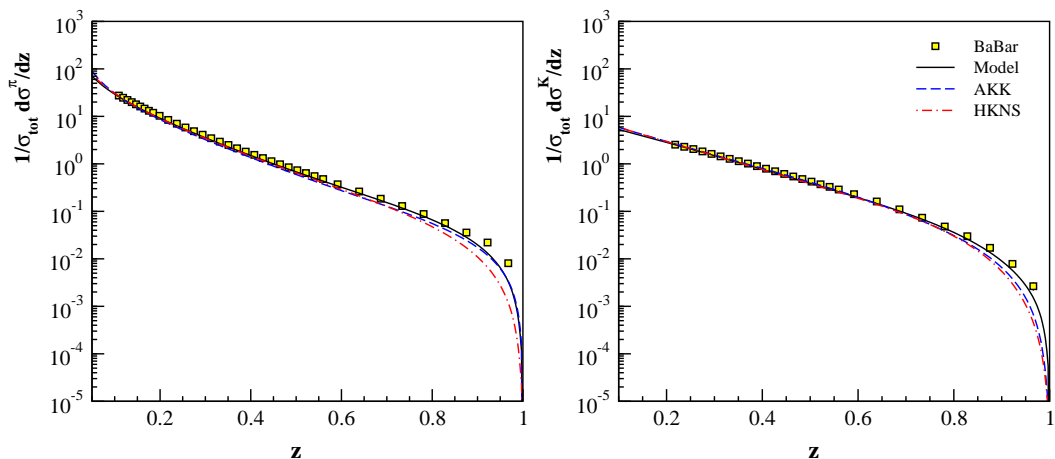}}
\vspace{-3cm}
\caption{As in Fig.~\ref{Belledata} but for {\it BABAR} data at $Q=10.54$~GeV~\cite{Lees:2013rqd}.}
\label{BaBardata}
\end{figure*}
\begin{figure*}[th]
\centerline{\includegraphics[width=0.7\textwidth,angle=-90]{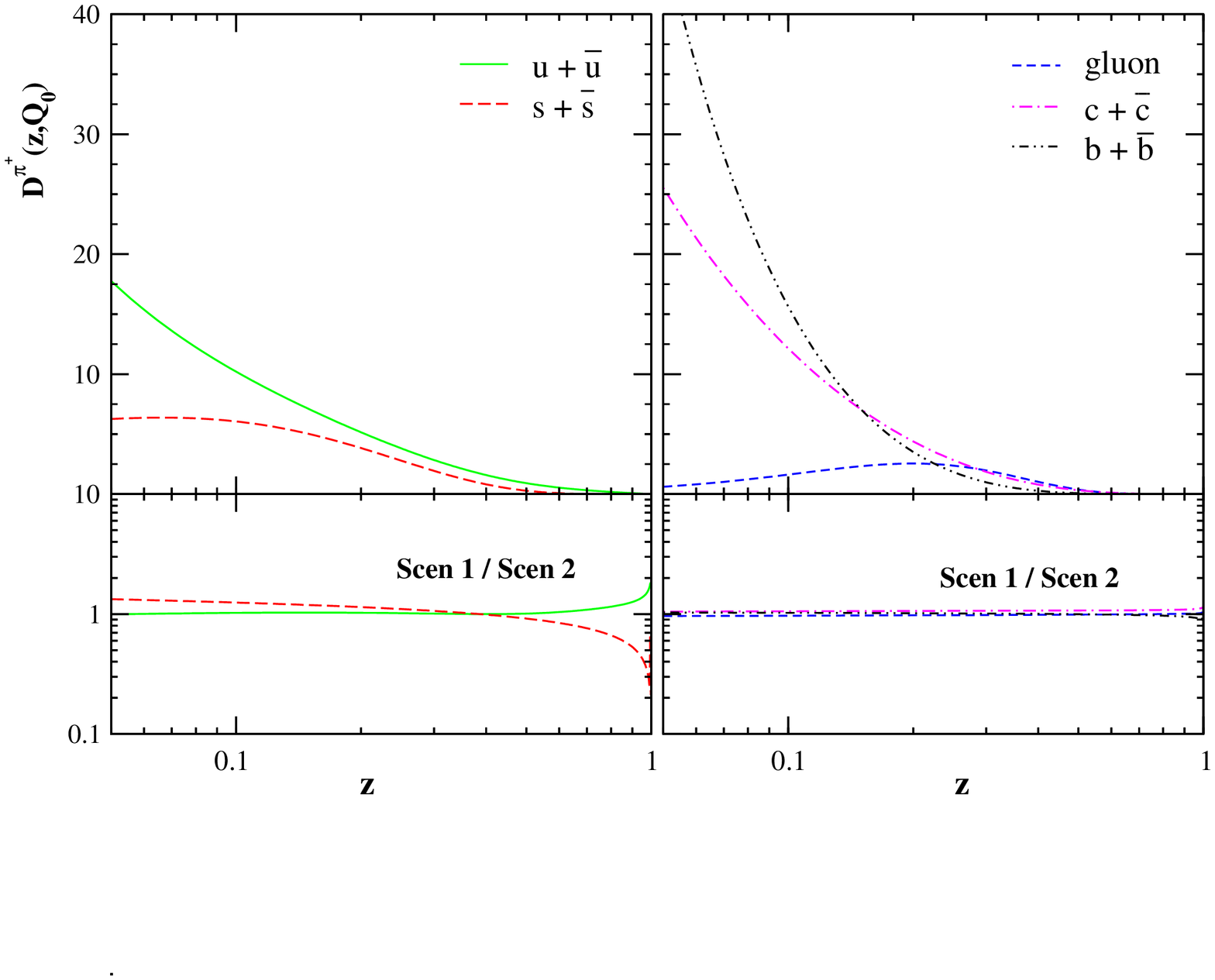}}
\vspace{-3cm}
\caption{Upper panels: NLO fragmentation functions for $\pi ^+$
at $Q_{0}^{2} = 1$ GeV$^2$, $m_{c}^{2}$ and $m_{b}^{2}$. Rest panels: ratios of our 
fragmentation functions from scenario 1 to the ones of scenario 2.}
\label{pFFsPLB}
\end{figure*}%
\begin{figure*}[th]
\centerline{\includegraphics[width=0.7\textwidth,angle=-90]{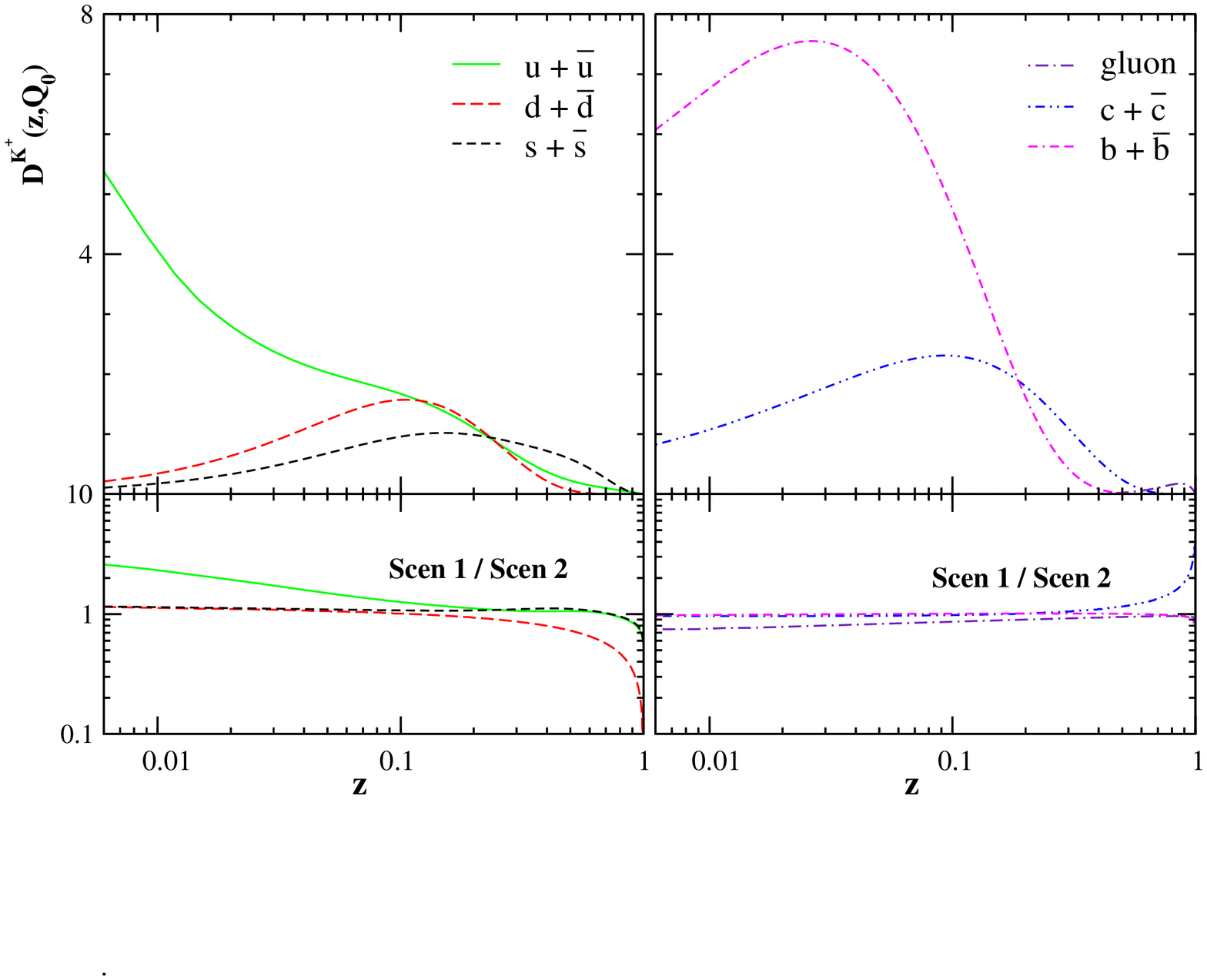}}
\vspace{-3cm}
\caption{Upper panels: fragmentation functions for $K^+$
at $Q_{0}^{2} = 1$ GeV$^2$, $m_{c}^{2}$ and $m_{b}^{2}$ at NLO. Rest panels: ratios of 
our fragmentation functions from scenario 1 to the ones of scenario 2.}
\label{KFFsPLB}
\end{figure*}

\section{The impact of Belle and {\it BABAR} data on FFs}
\label{sec3}
Recently the {\it Belle} \cite{Leitgab:2013qh} and {\it BABAR} \cite{Lees:2013rqd} Collaborations  published inclusive
hadron production cross sections at the c.m. energies of $10.52$~GeV and $10.54$~GeV, respectively.
These new data contain a purely $e^+e^- \rightarrow q\bar{q}$ sample, where
$q={u, d, s, c}$, since the c.m. energies are below the threshold of $b\bar{b}$ pair production.\\
The large amounts of data from {\it Belle} and {\it BABAR} Collaborations are available 
at the lower scales of $Q=10.52$~GeV and $Q=10.54$~GeV, while the energy scales of the other SIA experimental
data extracted from $29$ GeV to $91.2$ GeV and most of them are limited to results from experiments at LEP and SLAC at $Q=M_Z$.
In addition, these  new data include differential cross sections at larger $z$ values, i.e. $z > 0.7$.
Since the cross section measurements at the small $z$ depend on the
FFs at all larger $z$ values, the inclusion of these evaluations
will also lead to improved constraints on the FFs at
the $z$ values currently specified in global fits.\\
After adding these new data in analysis, our results for $\pi^\pm$ and $K^\pm$ at $Q=10.52$~GeV 
and $Q=10.54$~GeV are compared with experimental data in Figs~\ref{Belledata} and \ref{BaBardata}. 
Other FF models are also compared with the new data and these comparisons show a nice agreement between our model and these data.\\
In Figs.~\ref{pFFsPLB} and \ref{KFFsPLB}, we present the extracted NLO FFs of $\pi^+$ and $K^+$ in the initial scale $\mu_0$.
To show that adding these new data how much modify $\pi^+$ and $K^+$ FFs in our analysis, the ratio of obtained FFs by including
these new data (scenario 1) to ones without containing them (scenario 2) are also shown in Figs.~\ref{pFFsPLB} and \ref{KFFsPLB}.
According to these figures, the differences between these two scenarios are considerable at some regions of
$z$ for most of FFs and adding {\it Belle} and {\it BABAR} data change the light quark FFs more than the gluon and heavy quark FFs. 
Since the c.m. energies of {\it Belle} and {\it BABAR} data are below the threshold of $b$ 
quark production, it could be expected that the $b$ quark FFs do not change considerably. This is certified in the figures.

\section{Energy spectrum of the inclusive $\pi$ and $K$ in top quark decays}
\label{sec4}
In this section, we apply the extracted nonperturbative FFs to make our phenomenological predictions for the energy 
spectrum of the light mesons $\pi^\pm$ and $K^\pm$ produced through polarized top decays
\begin{eqnarray}\label{pros}
t(\uparrow)\rightarrow b+W^+ (g)\rightarrow \pi^\pm/K^\pm+X,
\end{eqnarray}
where $X$ stands for the unobserved final state.  Both the $b$-quark and the gluon may hadronize to 
the outgoing light mesons whereas the gluon contributes to the real radiation at NLO.\\
To obtain the energy distribution of the hadron $H$, we employ the factorization theorem of the 
QCD improved parton model where the energy distribution of a hadron can be expressed as the convolution 
of the nonperturbative FFs $D_i^H(z,\mu_F)$ with the parton-level spectrum as
\begin{equation}
\frac{d\Gamma}{dx_H}=\sum_{i=b,g}\int_{x_i^{min}}^{x_i^{max}}
\frac{dx_i}{x_i}\,\frac{d\Gamma_i^{pol}}{dx_i}(\mu_R,\mu_F)
D_i^H\left(\frac{x_H}{x_i},\mu_F\right).
\label{eq:master}
\end{equation}
Here, we define the scaled-energy fraction of hadron as $x_H=2E_{H}/(m_t^{2}-m_W^{2})$   and
$d\Gamma^{pol}_{i}/dx_i $ is the parton-level differential rates of the process
$t(\uparrow)\to i+W^+ (i=b,g)$. The analytical expressions for the parton-level differential decay widths
$d\Gamma_i^{pol}/dx_i$ at NLO are presented in Ref.~\cite{Nejad:2013fba}.
Here, the factorization and the renormalization scales are set to $\mu_R=\mu_F=m_t$.
\begin{figure*}[th]
\centerline{\includegraphics[width=0.7\textwidth,angle=-90]{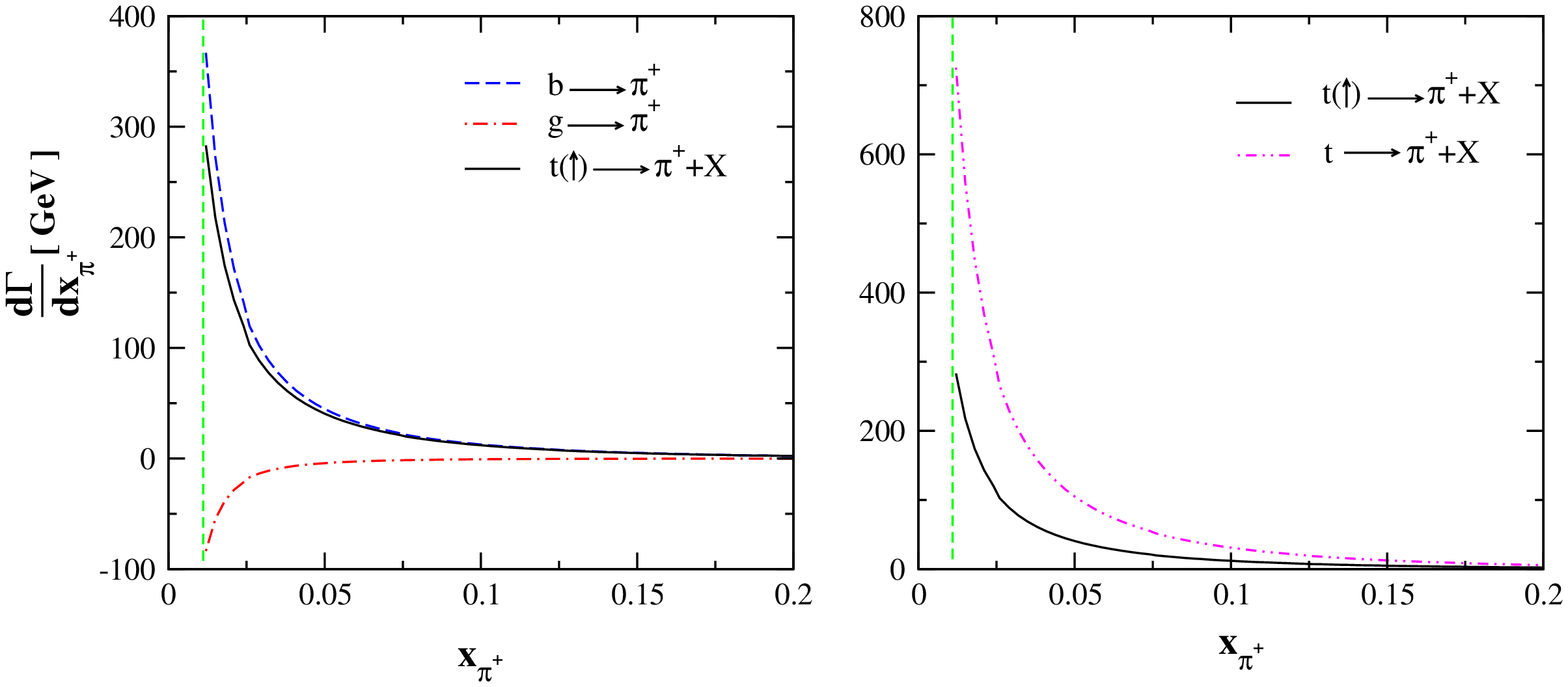}}
\vspace{-4.5cm}
\caption{$d\Gamma(t(\uparrow)\rightarrow \pi^++X)/dx_{\pi^{+}}$ as a function of $x_{\pi^{+}}$ (solid line)
 at $\mu_F=m_t$. Left panel: The NLO result is broken up into the
contributions due to $b\rightarrow B$ (dashed line) and $g\rightarrow B$ (dot-dashed line) fragmentation.
Right panel: The unpolarized (dot-dot-dashed line)
and polarized (solid line) partial decay rates at NLO.}
\label{Ppoltop}
\end{figure*}

\begin{figure*}[th]
\centerline{\includegraphics[width=0.7\textwidth,angle=-90]{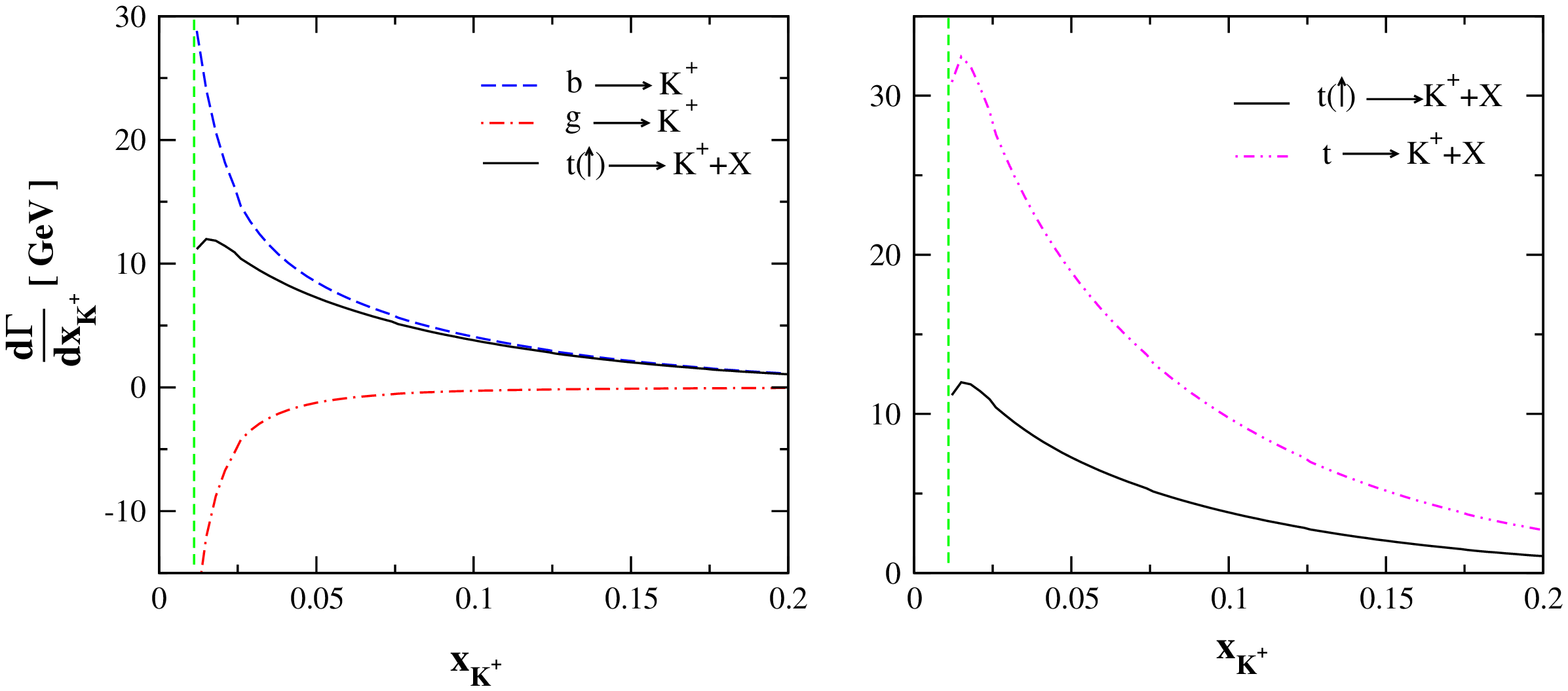}}
\vspace{-4.5cm}
\caption{As in Fig.~\ref{Ppoltop} but
 for $K^+$ at NLO.}
\label{Kpoltop}
\end{figure*}
In Figs. \ref{Ppoltop} and \ref{Kpoltop}, our predictions for the pion and kaon mesons are 
shown by studying the contributions of the $b\rightarrow \pi^+/K^+$ (dashed lines) and 
$g\rightarrow \pi^+/K^+$ (dot-dashed lines) fragmentation channels at NLO. 
As is seen, the gluon contribution is negative and appreciable only in the low $x_H$ region.
Note that the contribution of the gluon FF cannot be discriminated. It is
calculated to see where it contributes to $d\Gamma/dx_H (H=\pi^+, K^+)$. 
So, this part of the plot is of more theoretical relevance. In the
scaled energy of mesons as a experimental quantity, all
contributions including the b quark, gluon, and light quarks
contribute. The total contribution (solid line) at $\mu_F=m_t$ is presented too. 
In these figures, the scaled-energy ($x_H$) 
distribution of light mesons produced in unpolarized (dot-dot-dashed line) and 
polarized (solid line) top quark decays  are also studied. As is seen, in the unpolarized top decay the
partial decay width at the hadron level is higher than the one in the polarized top decay.

\section{Global minimization of $\chi^{2}$ and error calculation}
\label{sec5}
\begin{figure*}[th]
\centerline{\includegraphics[width=0.6\textwidth,angle=-90]{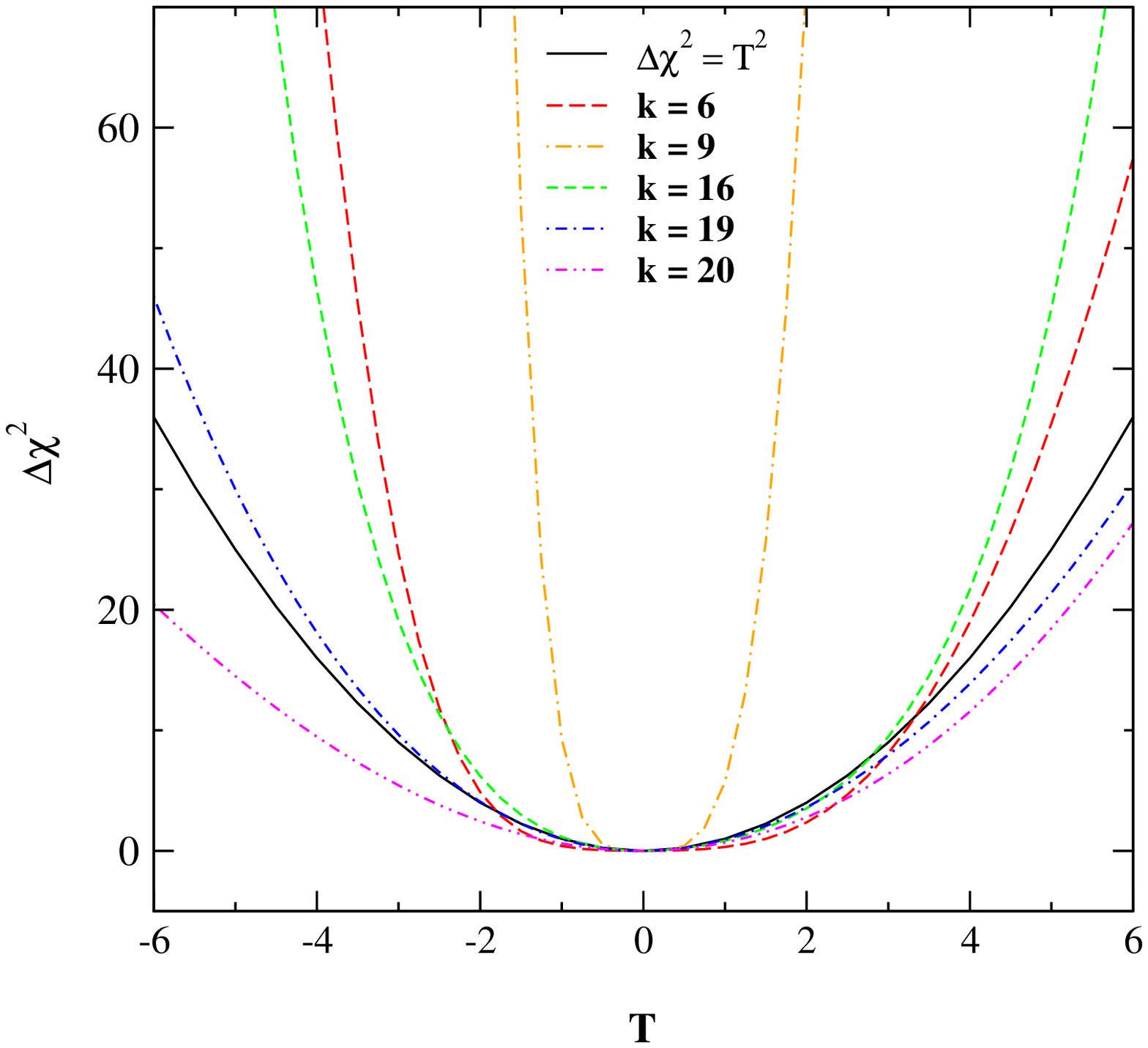}}
\vspace{-1cm}
\caption{Examples of pion $\Delta\chi^2$ deviations from
the expected quadratic behavior $\Delta\chi^2=T^2$ for
random sample eigenvector directions.}
\label{Deltachi2p}
\end{figure*}

\begin{figure*}[th]
\centerline{\includegraphics[width=0.6\textwidth,angle=-90]{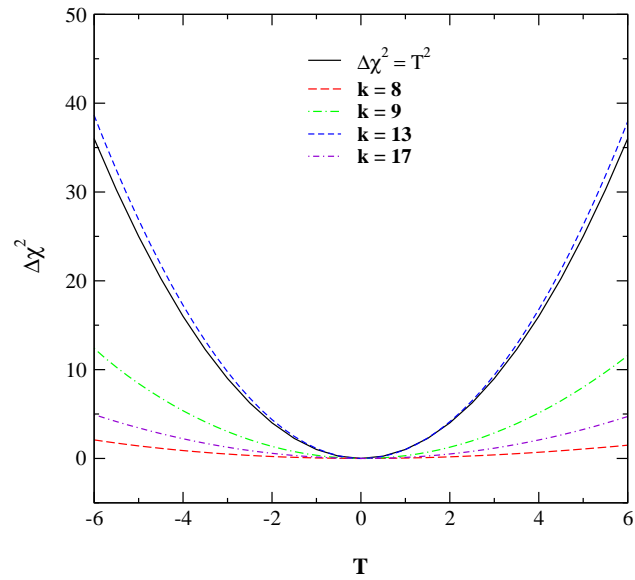}}
\vspace{-0.7cm}
\caption{As in Fig.~\ref{Deltachi2p} but for kaon, considering some random sample eigenvector directions.}
\label{Deltachi2k}
\end{figure*}

\begin{figure*}[th]
\centerline{\includegraphics[width=0.5\textwidth,angle=-90]{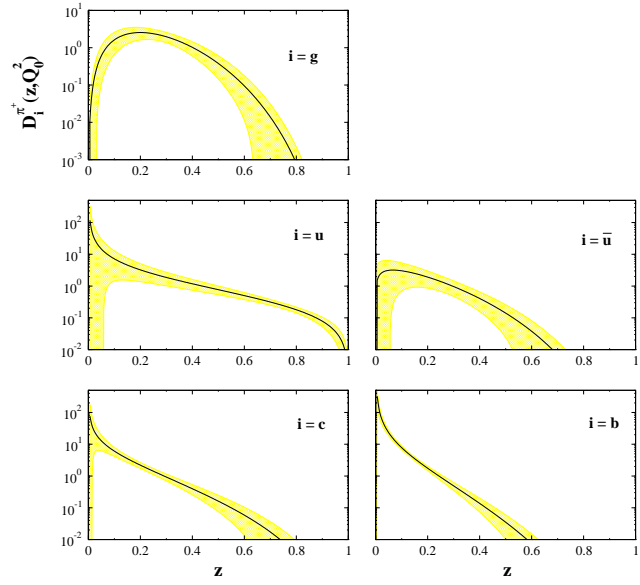}}
\caption{Fragmentation densities and their uncertainties
are shown for $\pi^+$ at
$Q_{0}^{2} = 1$ GeV$^2$, $m_{c}^{2}$ and $m_{b}^{2}$ at NLO.}
\label{errorpionNLO}
\end{figure*}

\begin{figure*}[th]
\centerline{\includegraphics[width=0.5\textwidth,angle=-90]{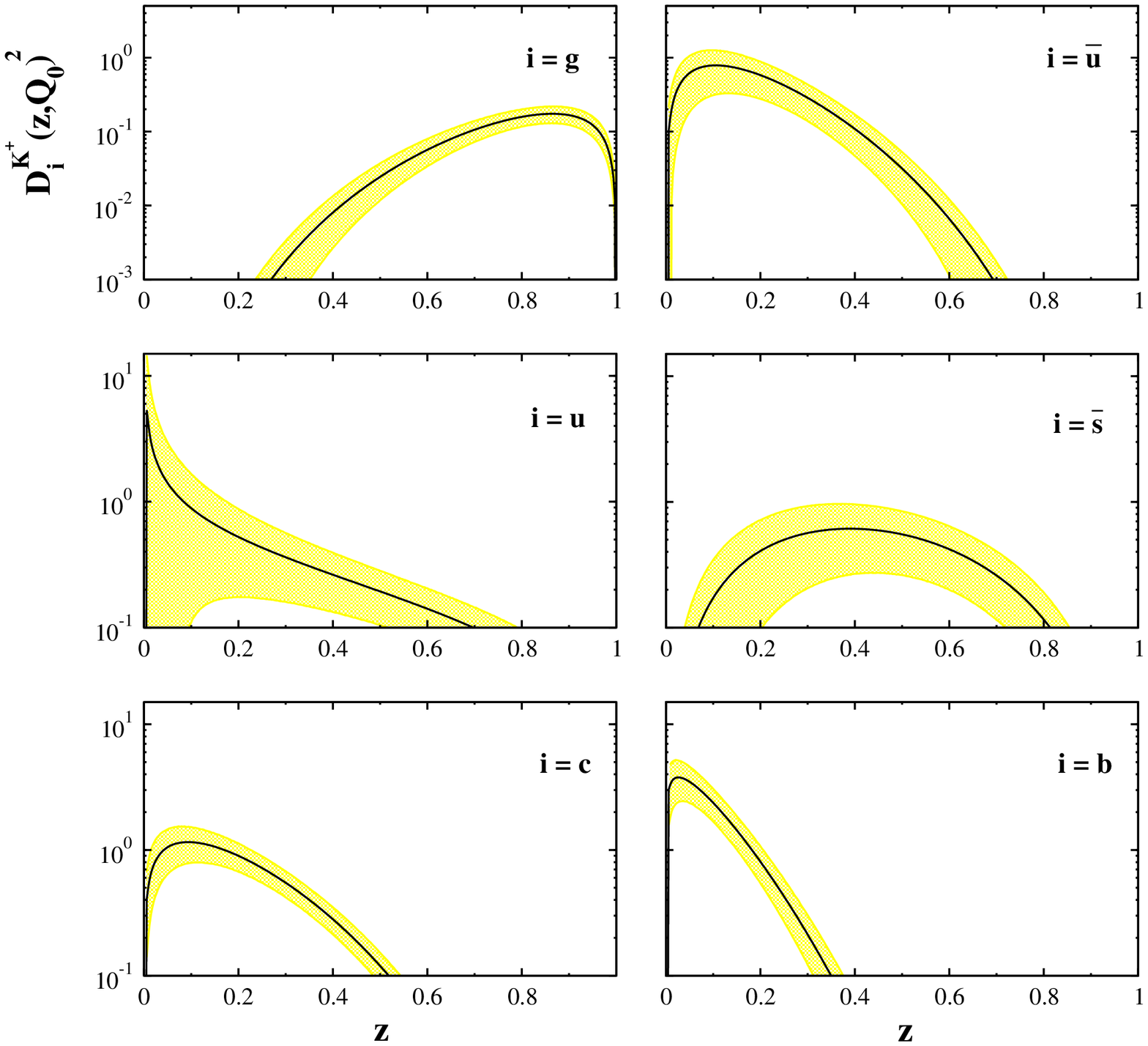}}
\caption{As in Fig.~\ref{errorpionNLO} but  fragmentation densities and their uncertainties
 for $K^+$ at NLO.}
\label{errorkaonNLO}
\end{figure*}

The free parameters in the proposed functional forms of  $\pi/K$ FFs
are determined by minimizing $\chi^2$ for differential cross section and asymmetry data 
in $x$ space. Results are reported in Tables.~\ref{tab:nlopionpara} and \ref{tab:nlokaonpara}. 
The $\chi^2$ for $k$ data points is defined as
\begin{equation}
\label{eq:chi2}
\chi^2=\sum_{j=1}^k (\frac{E_j-T_j}{\sigma^{E}_{j}})^2.
\end{equation}
Here,  $T_{j}$ and $E_j$ stand for the theoretical results and experimental values of data and $\sigma^{E}_{j}$
is the error of corresponding experimental value.
The obtained values of $\chi^2 /d.o.f$ for pion and kaon are $1.47$ and $1.54$ in our global fit, respectively.\\
After finding the appropriate parameters which minimize $\chi^2$, we can determine the behavior of
$\Delta \chi^2$ by moving away the parameters from their obtained values
\begin{equation} \label{eq:hessian}
\Delta\chi^2 \equiv \chi^2 - \chi_{\rm min}^2 = \sum_{i,j=1}^n H_{ij}(a_i-a_i^0)(a_j-a_j^0),
\end{equation}
where, $H_{ij}$ are the elements of the Hessian matrix 
$(H_{ij} = \left. \partial^2\,\chi^2/(2\partial a_i\partial a_j)\right|_{\rm min})$ and $n$ 
is the number of free parameters. According to the linear propagation of error, one can use the 
following formula for calculation of error on any quantity $F$
\begin{equation} \label{eq:Ferror}
(\Delta F)^2 =\Delta \chi^2{\sum_{j,k}^{n}\frac{\partial F}{\partial a_j}C_{jk}(a)\frac{\partial  F}{\partial a_k}}.
\end{equation}
Since Hessian matrix and its inverse ($C\equiv H^{-1}$ ), which is the error
matrix, are symmetric they contain a set of $n$ orthogonal eigenvectors $v_{ik}$ and eigenvalues $\lambda_{k}$
\begin{eqnarray}
\label{eq:eigeq}
\sum_{j=1}^n C_{ij}(a) v_{jk} &=& \lambda_k v_{ik}.
\end{eqnarray}
We can expanded the parameter variation around the global minimum in a basis of eigenvectors and
eigenvalues, that is,
\begin{equation} \label{eq:component}
\Delta a_{i}\equiv(a_i - a_i^0) = \sum_{k=1}^n e_{ik} z_k,
\end{equation}
where $e_{ik}\equiv \sqrt{\lambda_k}v_{ik}$. It can be shown that
the expansion of the $\chi^{2}$ in
the fit parameters ${a_{i}}$ near the global minimum (Eq.~\ref{eq:hessian}) reduces to
\begin{equation} \label{eq:hessiandiag}
\Delta \chi^2 = \sum_{k=1}^n z_k^2,
\end{equation}
where $\sum_{k=1}^n z_k^2\le T^2$ is the interior of a sphere of radius $T$.
In order to investigate that whether $\Delta\chi^{2}$ shows the assumed quadratic behavior of the 
parameters from the best fit, we present the dependence
of the global $\Delta\chi^{2}$  along some random samples of eigenvector
directions in Figs.~\ref{Deltachi2p} and \ref{Deltachi2k}.\\
To obtain the standard linear errors of FFs, we use
\begin{equation} \label{eq:heserror}
[\Delta D_{i}^{H}(z)]^2 =\Delta \chi^2{\sum_{j,k}^{n}\frac{\partial D_{i}^{H}(z,a_j)}{\partial a_j}C_{jk}(a)\frac{\partial
D_{i}^{H}(z,a_k)}{\partial a_k}},
\end{equation}
where $D_{i}^{H}(z;Q^2)$ is the evolved fragmentation density at $Q^2$ and $n$ is the number 
of parameters in the global fit. Finally we can compute the uncertainties of any FFs at 
any value of $Q^{2}$ by the QCD evolution. The $\pi^+$ and $K^+$ FFs and their uncertainties 
based on this method are presented in Figs.~\ref{errorpionNLO} and \ref{errorkaonNLO} at NLO.
More information and detailed discussions can be found in Refs.\cite{Hirai:2007cx,Pumplin:2001ct,Hirai:2003pm}.

%
%

\section{Conclusion and results}
\label{sec6}
In the present work we determined the nonperturbative FFs of partons into the pion and kaon
from global analysis on SIA and SIDIS data at NLO.
Our main aim was to show that adding the  
recent SIA data from {\it Belle} and {\it BABAR} Collaborations at $\sqrt{s}=10.52$ GeV and 
$\sqrt{s}=10.54$ GeV, respectively,  how much improve the results obtained for partonic FFs. 
Our analysis showed that these new data change the $(u, s)\rightarrow \pi^+$ FFs at the
large-$z$ region while the $s\rightarrow \pi^+$ FF is also changed at the low-$z$. 
As Fig. \ref{pFFsPLB} shows, these new data do not change the FFs of gluon and heavy quarks
into the pion.\\
Concerning the effects of new data on kaon FFs, as is seen from Fig.\ref{KFFsPLB},
the $u\rightarrow K^+$ FF is affected at low-$z$ ($z<0.2$) more than large-$z$, but the $d\rightarrow K^+$
FF affected at $z>0.07$. The FF of $g\rightarrow K^+$ is decreased  everywhere, e.g. about $25 \%$ at $z=0.01$.
The $c\rightarrow K^+$ FF is increased at large-$z$ when we consider new data. We hope our results can obvious
the effects of the recent new data on FFs for whom is studying the behavior of light meson FFs.\\
In \cite{Soleymaninia:2013cxa}, using the computed FFs we have studied the scaled-energy ($x_H$) distribution of the 
light mesons in unpolarized top quark decays and in the present work we made our predictions 
for the scaled-energy ($x_H$) distributions of the pion and kaon in polarized top decays.
The scaled-energy distribution of hadrons in polarized/unpolarized top quark decays at LHC enables us to deepen our
knowledge of the hadronization process. The universality and scaling violations of the pion and kaon FFs will be able to test
at LHC by comparing our NLO predictions with future measurements of $d\Gamma/dx_H$ and $d\Gamma(\uparrow)/dx_H$.\\
Note, the \texttt{FORTRAN} package containing our unpolarized fragmentation functions for pion and
kaon at LO and NLO can be obtained via e-mail from the authors.

\section{Acknowledgments}
\label{sec7}
We warmly acknowledge G. Corcella for valuable discussions, critical remarks and reading the manuscript.
A. N. K. and S. M. M. N. thank the CERN TH-PH division for its hospitality where a portion of this work was performed.
 We thank the School of Particles and Accelerators, Institute for Research in Fundamental Sciences (IPM) for financial support.



\begin{thebibliography}{}

\bibitem{Soleymaninia:2013cxa}
  M.~Soleymaninia, A.~N.~Khorramian, S.~M.~Moosavi Nejad and F.~Arbabifar,
  Phys.\ Rev.\ D {\bf 88}, 054019 (2013), arXiv:1306.1612 [hep-ph].

\bibitem{Kniehl:2012mn}
  B.~A.~Kniehl, G.~Kramer and S.~M.~Moosavi Nejad,
  Nucl.\ Phys.\ B {\bf 862} (2012) 720  [arXiv:1205.2528 [hep-ph]].  

\bibitem{Cacciari:2002re}
  M.~Cacciari, G.~Corcella and A.~D.~Mitov,
  JHEP {\bf 0212}, 015 (2002)
  [hep-ph/0209204].


\bibitem{Corcella:2001hz}
  G.~Corcella and A.~D.~Mitov,
  Nucl.\ Phys.\ B {\bf 623}, 247 (2002)
  [hep-ph/0110319].


\bibitem{Ma:1997yq}
  J.~P.~Ma,
 Nucl.\ Phys.\ B {\bf 506} (1997) 329.

\bibitem{Braaten:1993rw}
  E.~Braaten and T.~C.~Yuan,
Phys.\ Rev.\ Lett.\  {\bf 71} (1993) 1673.

\bibitem{Chang:1991bp}
  C.~-H.~Chang and Y.~-Q.~Chen,
 Phys.\ Lett.\ B {\bf 284} (1992) 127.  

\bibitem{Braaten:1993mp}
  E.~Braaten, K.~-m.~Cheung and T.~C.~Yuan,
  Phys.\ Rev.\ D {\bf 48} (1993) 4230.  


\bibitem{Scott:1978nz}
  D.~M.~Scott,
 Phys.\ Rev.\ D {\bf 18} (1978) 210.  

\bibitem{Bjorken:1977md}
  J.~D.~Bjorken,
  Phys.\ Rev.\ D {\bf 17} (1978) 171.  


\bibitem{Peterson:1982ak}
  C.~Peterson, D.~Schlatter, I.~Schmitt and P.~M.~Zerwas,
  Phys.\ Rev.\ D {\bf 27} (1983) 105.  

\bibitem{Suzuki:1977km}
  M.~Suzuki,
 Phys.\ Lett.\ B {\bf 71} (1977) 139.  



\bibitem{Amiri:1986zv}
  F.~Amiri and C.~-R.~Ji,
   Phys.\ Lett.\ B {\bf 195} (1987) 593.  

\bibitem{Nejad:2013vsa}
  S.~M.~M.~Nejad and A.~Armat,
  Eur.\ Phys.\ J.\ Plus {\bf 128} (2013) 121  [arXiv:1307.6351 [hep-ph]].  

\bibitem{Nejad:2014iba}
  S.~M.~M.~Nejad and D.~Mahdi,
  arXiv:1401.5223 [hep-ph].  


\bibitem{Leitgab:2013qh}
  M.~Leitgab {\it et al.}  [Belle Collaboration],
  Phys.\ Rev.\ Lett. {\bf 111}, 062002 (2013).

\bibitem{Lees:2013rqd}
  J.~P.~Lees {\it et al.}  [BaBar Collaboration],
  Phys.\ Rev.\ D {\bf 88}, 032011 (2013)
  [arXiv:1306.2895 [hep-ex]].


\bibitem{Albino:2008fy}
  S.~Albino, B.~A.~Kniehl and G.~Kramer,
  Nucl.\ Phys.\ B {\bf 803}, 42 (2008)  [arXiv:0803.2768 [hep-ph]].  

\bibitem{deFlorian:2007aj}
  D.~de Florian, R.~Sassot and M.~Stratmann,
  Phys.\ Rev.\ D {\bf 75}, 114010 (2007)
  [hep-ph/0703242 [HEP-PH]].
  
  
\bibitem{Hirai:2007cx}
  M.~Hirai, S.~Kumano, T.~-H.~Nagai and K.~Sudoh,
  Phys.\ Rev.\ D {\bf 75}, 094009 (2007)
  [hep-ph/0702250].
  
\bibitem{aleph91} D. Buskulic {\it et al.} (ALEPH collaboration),
                          Z. Phys. {\bf C66}, 355 (1995);
                  R. Barate {\it et al.}, Phys. Rep. {\bf 294}, 1 (1998).

 \bibitem{delphi91} P. Abreu {\it et al.} (DELPHI collaboration),
                          Eur. Phys. J. {\bf C5}, 585 (1998).
\bibitem{delphi91-2} P. Abreu {\it et al.} (DELPHI collaboration),
                          Nucl. Phys. {\bf B444}, 3 (1995).

\bibitem{opal91}  R. Akers {\it et al.} (OPAL collaboration),
                          Z. Phys. {\bf C63}, 181 (1994).

 \bibitem{sld91}   K. Abe {\it et al.} (SLD collaboration),
                          Phys. Rev. {\bf D69}, 072003 (2004).


 \bibitem{tasso34_44} W. Braunschweig {\it et al.} (TASSO collaboration),
                          Z. Phys. {\bf C42}, 189 (1989).
\bibitem{tpc29}      H. Aihara {\it et al.} (TPC collaboration),
                          Phys. Rev. Lett. {\bf 52}, 577 (1984);
                                           {\bf 61}, 1263 (1988).
 
\bibitem{topaz58} R. Itoh {\it et al.} (TOPAZ collaboration),
                          Phys. Lett. {\bf B345}, 335 (1995). 
                                           
\bibitem{dglap}
  V.~N.~Gribov and L.~N.~Lipatov,
  Sov.\ J.\ Nucl.\ Phys.\  {\bf 15}, 438 (1972)
  [Yad.\ Fiz.\  {\bf 15}, 781 (1972)];
  G.~Altarelli and G.~Parisi,
  Nucl.\ Phys.\ {\bf B126}, 298 (1977);
  Yu.~L.~Dokshitzer,
  Sov.\ Phys.\ JETP {\bf 46}, 641 (1977)
  [Zh.\ Eksp.\ Teor.\ Fiz.\  {\bf 73}, 1216 (1977)].

\bibitem{Nejad:2013fba}
  S.~M.~M.~Nejad,
  Phys.\ Rev.\ D {\bf 88} (2013) 094011  [arXiv:1310.5686 [hep-ph]].  
   
\bibitem{Kretzer:2000yf}
  S.~Kretzer,
  Phys.\ Rev.\  D {\bf 62}, 054001 (2000)
  [arXiv:hep-ph/0003177].

\bibitem{Kniehl:2005de}
  B.~A.~Kniehl and G.~Kramer,
  Phys.\ Rev.\  D {\bf 71}, 094013 (2005)
  [arXiv:hep-ph/0504058].
\bibitem{Binnewies:1994ju}
  J.~Binnewies, B.~A.~Kniehl and G.~Kramer,
  Z.\ Phys.\  C {\bf 65}, 471 (1995)
  [arXiv:hep-ph/9407347].



\bibitem{Kneesch:2007ey}
  T.~Kneesch, B.~A.~Kniehl, G.~Kramer and I.~Schienbein,
  Nucl.\ Phys.\ B {\bf 799}, 34 (2008)  [arXiv:0712.0481 [hep-ph]].  




\bibitem{Kniehl:2008zza} 
  B.~A.~Kniehl, G.~Kramer, I.~Schienbein and H.~Spiesberger,
  Phys.\ Rev.\ D {\bf 77}, 014011 (2008).  


\bibitem{Webber:1983if}
  B.~R.~Webber,
   Nucl.\ Phys.\ B {\bf 238} (1984) 492.  

\bibitem{Hermes05}
  A.~Airapetian {\it et al.}  [HERMES Collaboration],
  Phys.\ Rev.\  D {\bf 71} (2005) 012003
  [arXiv:hep-ex/0407032].


\bibitem{Alekseev:2009ac}
  M.~Alekseev {\it et al.}  [COMPASS Collaboration],
  Phys.\ Lett.\ B {\bf 680}, 217 (2009)
  [arXiv:0905.2828 [hep-ex]].



\bibitem{Alekseev:2010ub}
  M.~G.~Alekseev {\it et al.}  [COMPASS Collaboration],
  Phys.\ Lett.\ B {\bf 693}, 227 (2010)
  [arXiv:1007.4061 [hep-ex]].


\bibitem{Pumplin:2001ct}
  J.~Pumplin, D.~Stump, R.~Brock, D.~Casey, J.~Huston, J.~Kalk, H.~L.~Lai and W.~K.~Tung,
  Phys.\ Rev.\ D {\bf 65}, 014013 (2001)
  [hep-ph/0101032].

\bibitem{Hirai:2003pm}
  M.~Hirai {\it et al.}  [Asymmetry Analysis Collaboration],
  Phys.\ Rev.\ D {\bf 69}, 054021 (2004)
  [hep-ph/0312112].



\end{thebibliography}
\end{document}